\documentclass[%
 reprint,
 superscriptaddress,
 amsmath,amssymb,
 aps,
 pra,
]{revtex4-1}

\usepackage[pdftex]{graphicx}% Include figure files
\usepackage[pdftex]{hyperref}
\usepackage[normalem]{ulem}
\usepackage{dcolumn}% Align table columns on decimal point
\usepackage{bm}% bold math
\usepackage{color}
\usepackage{gensymb}
\usepackage{hyperref}

\newlength{\mywidth}
\usepackage[normalem]{ulem}

\begin{document}

\title{Vortex creation, annihilation and nonlinear dynamics in atomic vapors}% Force line breaks with \\

\author{Pierre Azam}%
\affiliation{Institut de Physique de Nice, Universit\'e C\^ote d'Azur, CNRS, Nice, France}%
\author{Adam Griffin}
\affiliation{Institut de Physique de Nice, Universit\'e C\^ote d'Azur, CNRS, Nice, France}
\author{Sergey Nazarenko}%
\affiliation{Institut de Physique de Nice, Universit\'e C\^ote d'Azur, CNRS, Nice, France}%
\author{Robin Kaiser}%
\affiliation{Institut de Physique de Nice, Universit\'e C\^ote d'Azur, CNRS, Nice, France}%

\date{\today}% It is always \today, today,

\begin{abstract}

%We exploit two new techniques for generating vortices and controlling their subsequent interactions in an optical beam passing through a nonlinear atomic vapor. The two connected mechanisms, the snake instability and two-beam interference, allow precise control of the vortex positions which in turn allows us to observe strong interactions leading to vortex annihilations. With this improved controlled nonlinear system, we get closer to the pure hydrodynamic regime than previous experiments which is important in the context of vortex studies. The use of a specific wavefront sensor offers a rapid and precise tool for vortex detection and the study of their dynamics by giving us a direct access to the fluid's density and velocity. Finally, we developed a relative phase shift method which mimics a time evolution process without the need to change non-linear parameters, resulting in improved interpretability in the context of fluid dynamics. The observed accelerated vortex interactions constitute an important step toward the experimental implementation of 2D turbulent state consisting of many chaotically interacting vortices.

We exploit new techniques for generating vortices and controlling their interactions in an optical beam in a nonlinear atomic vapor. A precise control of the vortex positions allows us to observe strong interactions leading to vortex dynamics  involving annihilations. With this improved controlled nonlinear system, we get closer to the pure hydrodynamic regime than in previous experiments while a wavefront sensor offers us a direct access to the fluid's density and velocity. Finally, we developed a relative phase shift method which mimics a time evolution process without changing non-linear parameters. These observations are an important step toward the experimental implementation of 2D turbulent state.

\end{abstract}

\maketitle

\section{Introduction}

Fluid-like properties of light in nonlinear Kerr-like media is an interesting and rapidly developing subject of research. In particular, they were investigated in photorefractive crystals \cite{Wan07, Michel18} and thermo-optic media \cite{Vocke15, Vocke16}. In the present paper, we will discuss light propagation in atomic vapors which was proven an effective alternative platform for fluids of light studies. In the past, this experimental platform was used to implement vortex creation \cite{,swartzlander_optical_1992, tikhonenko_observation_1996}, photon pre-condensation \cite{Santic18} and to measure the Bogoliubov dispersion relation \cite{Fontaine18, Fontaine20}.

%Turbulence is the most common type of fluid flow in nature.
There has been a vast amount of studies devoted to vortex interactions and turbulence, yet their theory is far from being complete. In the context of optical fluids, there have been significant advances in studying turbulence theoretically \cite{DYACHENKO,Nazarenko11} and numerically \cite{onorato,proment,Griffin20}.
%Optical fluids are compressible and acoustic waves play an important role in optical turbulence. Even in absence of vortices, there could be a nontrivial turbulent state with a large number of random mutually interacting sound waves: this is so-called acoustic wave turbulence \cite{Nazarenko11}.
Furthermore, with optical vortices having quantized circulations, optical turbulence is similar to turbulence in superfluids (e.g. liquid Helium) and Bose-Einstein condensates \cite{abid03,henn09}.

Optical turbulence was previously implemented in 1D systems in liquid crystals \cite{Bortolozzo,laurie} and optical fibers \cite{turitsyn}. However, no experimental implementation of 2D optical turbulence has been done so far, which is related to experimental challenges of minimizing the dissipation to nonlinearity ratio and finding optimal experimental setups in which a large number of vortices could be created and maintained in the system for a sufficiently large period of effective time so that random hydrodynamic motion of vortices could lead to universal turbulent statistics.

In the present paper, we study strongly nonlinear multiple-vortex generation of optical vortices in atomic vapors and their interactions through time-like evolution. The techniques demonstrated in this work could be scaled up to systems with large numbers of randomly moving strongly nonlinear vortices with hydrodynamic properties thereby implementing turbulent states. The strategy to achieve these results are based on a proper choice of the nonlinear parameters and the initial configuration of the incident beam, providing a robust and efficient mechanism for vortex generation. We developed a specific method precisely controlling evolution time of processes occurring in our fluid which is equivalent to the Taylor's frozen turbulence hypothesis \cite{Taylor38}, and which allows us to observe more evolved vortex creations and interactions at shorter effective times.

An interesting mechanism of vortex generation is the so-called snake instability. The snake instability of 1D solitons was first discovered in the context of 2D dispersive compressible fluids arising in plasma context \cite{KP1970}, it was further studied in \cite{kuznetsov_instability_1988} and later generalized to nonlinear defocusing media in \cite{kuznetsov_instability_1995}. 
In the context of defocusing nonlinear optics (and Bose-Einstein condensates) solitons are dark (the light intensity is less inside the soliton than in the ambient medium). In the case of optical media with saturation-type nonlinearity, such soliton solutions  were treated theoretically in \cite{krolikowski_dark_1993,krolikowski_dark_1994}.
In this context, the snake instability  is known to be a precursor to generation of two-dimensional dark solitons  \cite{jones_motions_1982} and vortex nucleation \cite{law_optical_1993}. The nucleation of vortices via the snake instability \cite{maitre2021} and their subsequent dynamics \cite{Boulier15} have also been studied in polaritons.
Dark solitons and their instability in two-dimensional condensates leading to the creation of vortices were recently studied numerically in \cite{verma_dark_2017, Griffin20}. In the latter paper, trains of multiple dark solitons were created in a Josephson-Junction setup where  two cavities with initially un-equal densities are separated by a potential barrier. This was shown to be an ideal setup for generating turbulence because the instability of multiple solitons leads to the creation of a large number of well sustained vortices. 

%A closely related mechanism of turbulence excitation was suggested in \cite{rodrigues_turbulence_2020} using  counter-streaming superfluids of light which, at an intermediate step, also leads to soliton wave trains which further undergo instability resulting in vortex nucleation.

In the present paper, we report on experiments using the snake instability and similar mechanisms to create vortices with hydrodynamic properties. This work can be viewed as a stepping stone to building future experiments on 2D optical turbulence. Initial efforts on optical vortices created as a result of the development of instability of dark soliton stripes were previously reported in atomic Rubidium experiment \cite{tikhonenko_observation_1996} and in photorefractive crystals \cite{mamaev_propagation_1996}.
The novel features exploited in the present paper include:

1. In contrast to a holographic technique \cite{Verrier11, Vocke16, Michel20}, we use  a lateral shearing interferometer \cite{Chanteloup05, Primot95, Velghe05}: a 2D grating create several identical replicas of the incoming wave front that interfere with each other. A real-time analysis in the Fourier space gives us access to the local intensity profile and phase gradients of the observed wave-front. This lateral shearing interferometry technique has been  used for studying aberrations in regimes of short wavelengths \cite{Chanteloup05,Primot95} or intense beams and in the context of adaptative optics and ophthalmology. 
In the context of vortices in nonlinear media, this imaging of the output field at the exit facet of the nonlinear medium allows us to accurately identify optical vortices and track their precise location wihtout using regular interferometers requiring a reference beam, a scanning process and a post-analysis.

2. A better control of beam’s initial conditions. Using specific geometries to optimize vortex nucleation leading to an acceleration of vortex creation process and even to controlled vortex interactions. On another hand, increasing the background beam size and shortening the non-linear length (e.g. compared to \cite{tikhonenko_observation_1996}) reduces the impact of  dispersion on the vortex motion. As presented in our previous study \cite{azam21}, we can approximate the background on which the vortices rest as flat, thereby minimizing  the impact the local  refraction index fluctuations onto the vortex motions. This allows us to assume that the hydrodynamic motion of vortices is the main observed process.
%Saturable nonlinear medium \cite{krolikowski_dark_1993,,tikhonenko_vortex_1998}\\
%Optical vortex solitons and the stability of dark soliton stripes \cite{law_optical_1993}\\
%Optical Vortex solitons Observed in Kerr Nonlinear Media \cite{swartzlander_optical_1992}\\

\section{Model \& Experimental setup}

The  system studied here consists of a (2+1) dimension optical field propagating along $z$ through a nonlinear medium. In the paraxial approximation, the slowly varying complex amplitude $\psi({\bm r},z)$ of the laser field can be described by the nonlinear Schr\"odinger equation:

  \begin{eqnarray}
i \frac{\partial \psi }{\partial z}=\left( -\frac{\alpha}{2}\nabla_{\bf r}^{2} - k_0\Delta n -i\frac{\eta}{2} \right ) \psi\label{gpe}
  \end{eqnarray}
  where ${\bf r}=(x,y)$ is the coordinate in the transverse to the beam plane, $z\approx tc$ (where $t$ is time and $c$ the speed of light) is the distance along the beam,
 $\alpha=1/k_0$ is the kinetic energy (dispersion) parameter, $k_0=2\pi/\lambda_0$ corresponds to the wave vector of the laser beam ($\lambda_0$ being its wavelength in vacuum), and the beam intensity is $I=n_0\epsilon_0c\left|\psi({\bm r},z) \right|^{2}/2$, with $n_0$ being the linear index of refraction and $\epsilon_0$---the vacuum permittivity. Further, $\Delta n$ is the nonlinear refractive index and the parameter $\eta$, proportional to the inverse of the absorption length, quantifies the linear absorption
rate of the medium.
% and is define as $|\psi({\bm r},z)|^2=|\psi({\bm r},0)|^2\exp(\eta z/2)$.

%\textcolor{red}{Also should the prop in (2) be equal? (AG)\\ There is a prop because it is time a constant depending on the number of atom, etc... (PA)\\ }\\
In our experiment,  the nonlinear refractive index is given by a saturating nonlinearity model:
  \begin{eqnarray}
\Delta n\propto\frac{\delta}{1+4\delta^2/\Gamma^2}\frac{I/I_{sat}}{1+4\delta^2/\Gamma^2+I/I_{sat}}\label{dn}
  \end{eqnarray}
where $\delta$ is the detuning between the laser and the atomic resonance with respect to the $^{87}$Rb D2 transition $5S_{1/2}(F=2)- 5P_{3/2}(F=1,2,3)$, $\Gamma=2\pi\cdot 6.06$MHz is its natural line width and $I_{sat}$ is the saturation intensity. For negative nonlinear refractive index ($\Delta n<0$), the medium will have a defocusing response that corresponds to an effective repulsive photon-photon interaction.
Two typical lengths in the system are important: the nonlinear length scale $z_{NL}=1/(k_0|\Delta n|)$ corresponding to the effective time scale of the propagation and  the healing length $\Lambda=\sqrt{z_{NL}/(2k_0)}$, corresponding to the minimal length scale for density variation in the transverse plane.

\begin{figure}[!htb]
\center{
  \includegraphics[width=0.80\linewidth]{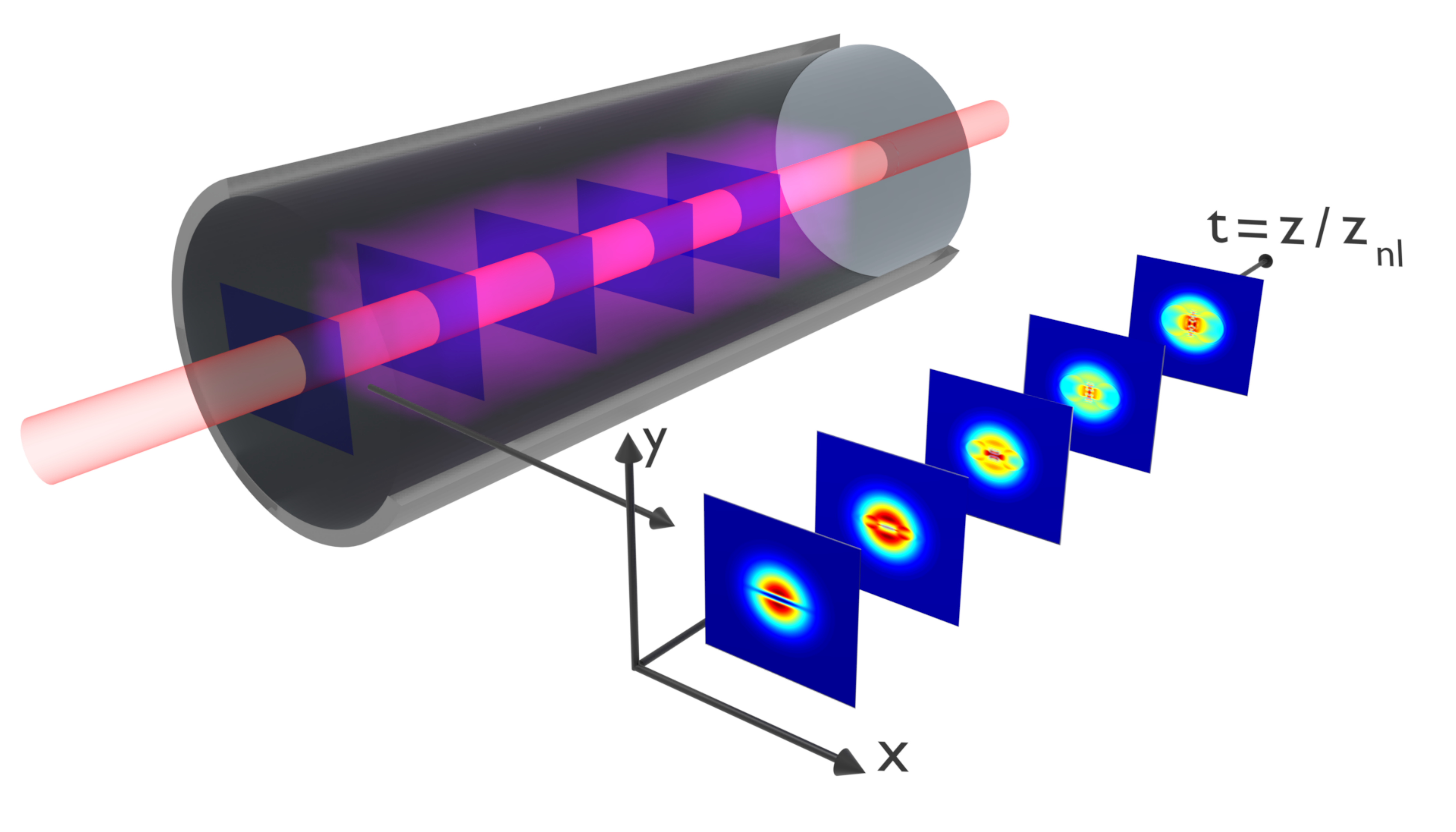}
\caption{Experimental scheme: successive images show how the system evolves from the engeneered intial condition after propagating along the re-normalized effective time $t\propto z/z_{NL}$}
\label{Fig1}}
\end{figure}

Using the Madelung transformation  $\psi({\bm r},z)=\sqrt{\rho({\bf r},z)}\exp(i\phi({\bf r},z))$, we can write a system of hydrodynamical equations for the electric field:
\begin{eqnarray}
&\partial_z\rho+\nabla_{{\bf r}} \cdot(\rho{\bf v})=-\eta \rho, \label{Hydro1}
\\
&\partial_z{\bf v}+({\bf v} \cdot \nabla_{{\bf r}})  {\bf v}=\nabla_{{\bf r}} h \label{Hydro2} .
\end{eqnarray}
This formulation describes the laser beam as a fluid of density $\rho$ which flows at a velocity ${\bf v}=\frac{1}{k_0}\nabla{\bf \phi}$ in the transverse to the beam propagation plane. 
Here, the fluid's specific enthalpy is defined as
$$
h(\rho)= \Delta n(\rho) + \frac 1 {2k_0^2} \frac{\nabla^2 \sqrt \rho}{\sqrt \rho},
$$
where the second term is called ``quantum enthalpy" (since this term is absent in the classical fluid equations).
Our fluid of light can be characterized by the speed of sound in the medium $c_s=\sqrt{\rho \frac{ \partial (-\Delta n)}{
\partial \rho}}$.
%. = c\sqrt{|\Delta n|}$.
%Note that we will deal with the defocusing case so that $\Delta n <0$.
\\

%\section{Experimental setup}

The optical field is composed of a gaussian background beam (with waist $w_{G}=1.1$mm and power $P_{G}=800$mW) overlapped with an elliptical gaussian beam (with dimensions $w_{x}=730\mu$m and $w_{y}=140\mu$m) having an equal central intensity as the background beam. For desctructive interference (relative phase $\phi=\pi$) we obtain close to zero intensity in the middle of the elliptical beam (as in the first image on Fig.~\ref{Fig1}).

This field propagates along the $z$-axis into a cylindrical cell of length $L=7$cm and diameter $2.5$cm filled with a natural isotopic mixture of $^{85}$Rb and $^{87}$Rb. Such a vapor behaves like a nonlinear medium whose strength can be tuned experimentally. The nonlinear medium has a finite size, and reducing $z_{NL}$ through $\Delta n$ mimics an increase of the effective propagated time. The nonlinearity is measured using the nonlinear phase shift $\Phi_{NL}=L/z_{NL}$.
It can be adjusted through the beam intensity $I$, the detuning $\delta$ or the atomic density of the vapor $\rho_{at}$ (tunable with the vapor temperature)\cite{Zhang15}. 
Typical experimental parameters are: the background intensity  $I=4\times10^5$ W/m$^2$, the atomic density of $\rho_{at}\approx2\times10^{19}$ atoms/m$^{3}$ (at a temperature of $T\approx 120$ \degree C), and the laser detuning scanned from $-10$ GHz to $-1$ GHz. The temperature measurements method is explained in the appendix.\\
By progressively increasing the nonlinearity using the detuning (i.e. reducing $z_{NL}$),   we observe the fluid evolution after different effective times at the output facet as shown in Fig.~\ref{Fig1}. A Phasics wave-front sensor captures the near field at the output of the medium (\cite{Primot95,Velghe05,Chanteloup05}). This  offers the possibility to  measure the intensity and phase of the field directly, which gives   the optical fluid density and velocity   in our experiment.

\section{Initial velocity}

\begin{figure}[!htb]
\center{
  \includegraphics[width=1\linewidth]{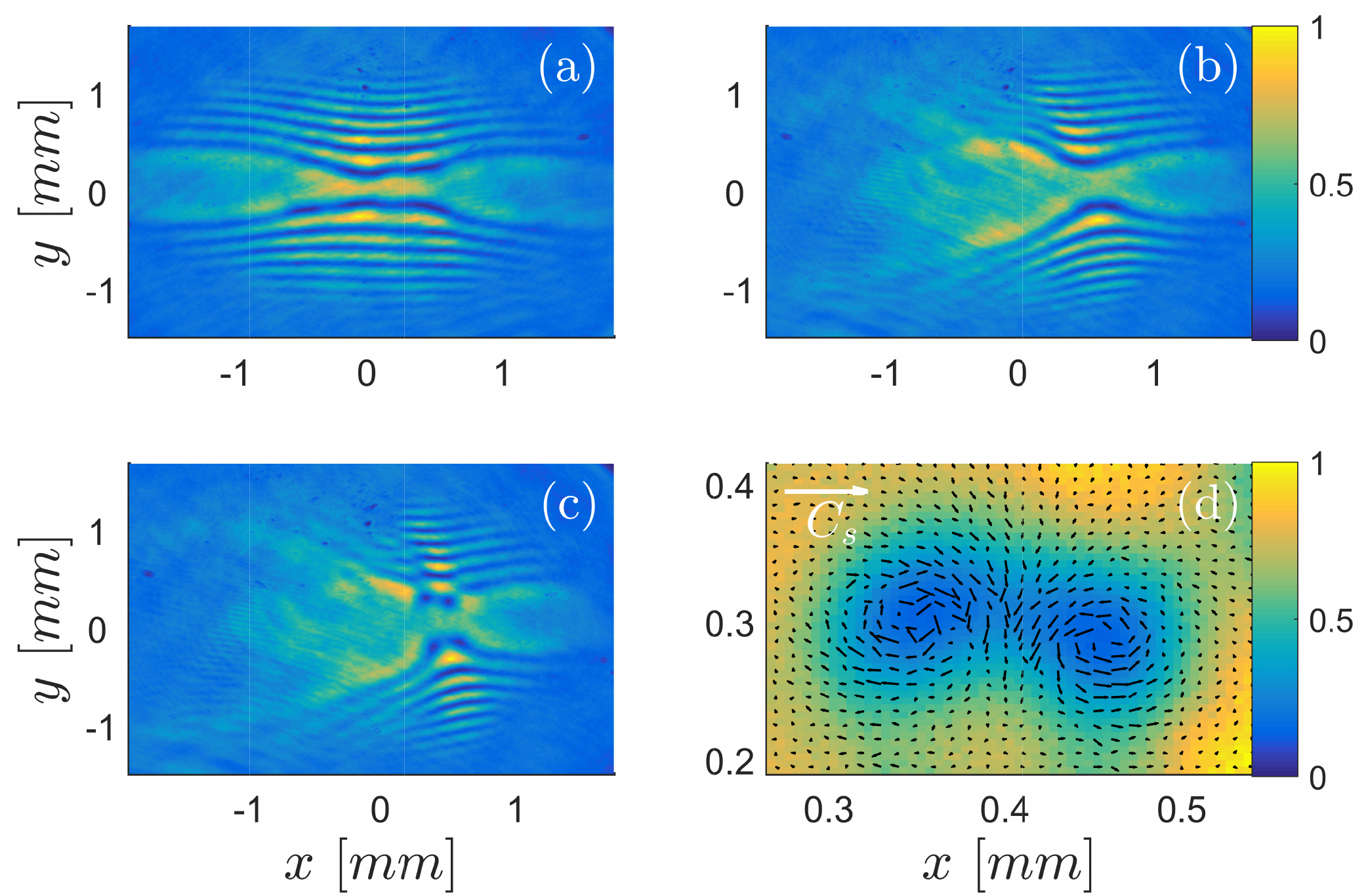}
\caption{Experimental observation of vortex creation by increasing the relative angle in the xz plane between beams (a) $\theta=0$, (b) $4.4\times 10^{-4}$ rad, (c) $5.2\times 10^{-4}$ rad and
(d) is a zoomed vortex pair with fluid density (colormap) and velocity (vectors, normalized by $c_s$ shown as the white arrow).}
\label{Fig2}}
\end{figure}

Due to  the action of the refractive index  when entering the nonlinear medium, the initially elliptical dark stripe decays into a symmetric set of  dark solitons on which snakelike bendings appear. Similar soliton trains appear as a result of decay of initial density discontinuities e.g. in the Josephson junction setup \cite{Griffin20}.  As we can see Fig.~\ref{Fig2} (a), the  bending pattern is concave because the initial light intensity is not strictly one dimensional:  the reference beam intensity and the dark stripe depletion are larger at the beam center.
%It is oriented towards the beam center where nonlinearity increases as we can see Fig.~\ref{Fig2} (a).
%dark soliton stripes. When increasing the nonlinearity, snakelike bendings appear on each stripe oriented towards the beam center where the intensity is higher. Sound excitations are also visible on the outer sides of the solitons (Fig.~\ref{Fig2} (a) as a typical example).
Such transverse bending and amplitude modulation serve as an initial seed for the snaking instability which  distorts the soliton stripes further leading to their breakup \cite{kuznetsov_instability_1988}.
At $\delta=-1$ GHz corresponding to $|\Delta n|=8.8\times 10^{-5}$ (at the end of the nonlinear medium in the central part of the beam) the system evolves far enough to observe well developed snake instability bendings. These parameters correspond to a nonlinear phase shift $\Phi_{NL}=50$.

\begin{figure}[!htb]
\center{
  \includegraphics[width=1\linewidth]{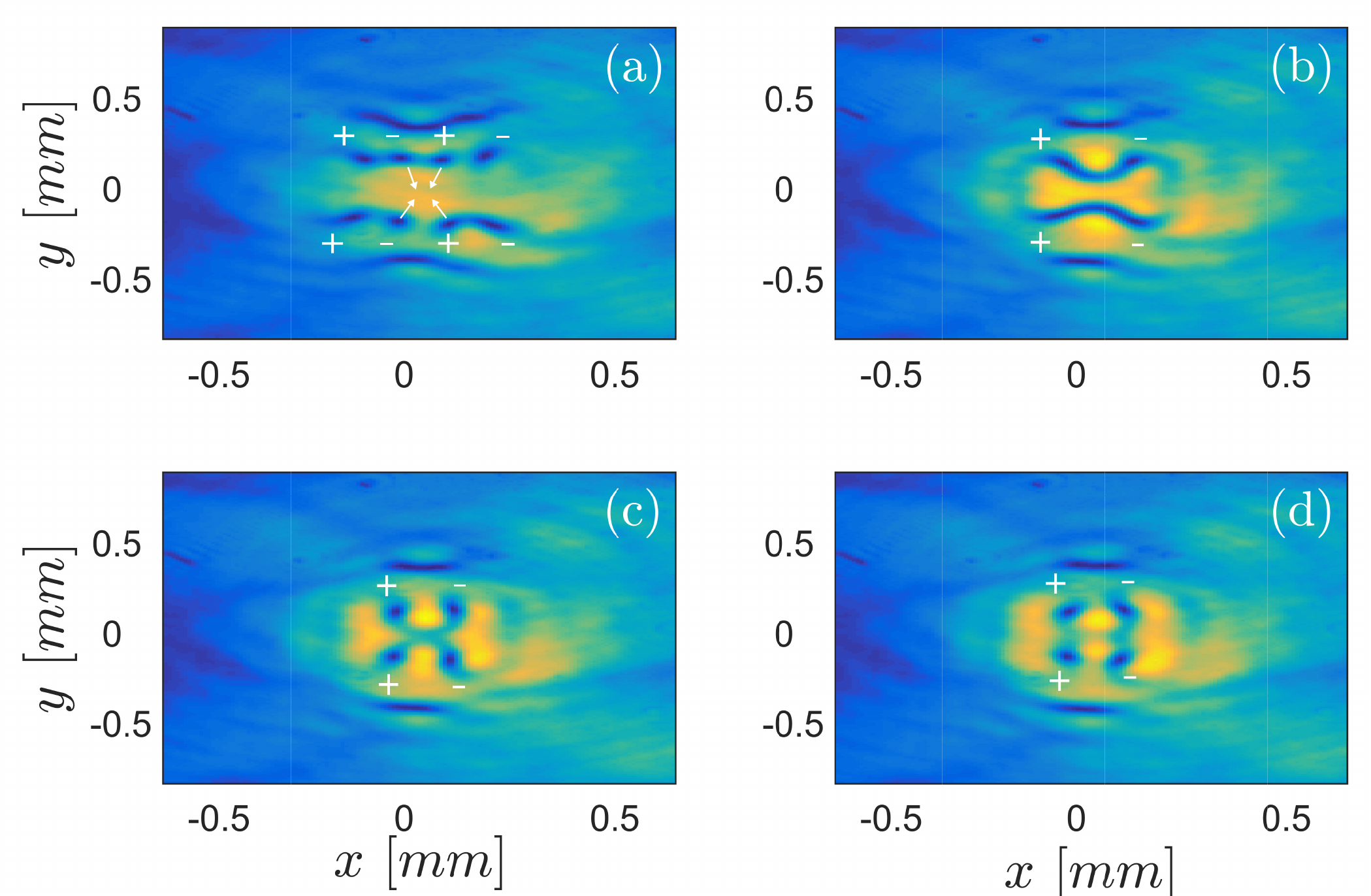}
\caption{Experimental observation of  creation and interaction of vortices: We are using an initial convergent elliptical beam which is shifted by phase $\phi$ relative to the background beam a) to d):  $\varphi=0.96\pi, 1\pi, 1.05\pi, 1.10\pi$  respectively.}
\label{Fig3}}
\end{figure}

As shown in Fig.~\ref{Fig2} (a), in a simple configuration snake instabilities do not have the necessary evolution time  to break into vortices. However, increasing  the relative initial velocity between the background and the dark stripe results in an acceleration of the snake instability growth and leads to  vortex nucleation. Such a  relative transverse velocity can be expressed as $v=r/t\approx\theta c$ where the angle between the background and the elliptical beam in the xz plane is approximated by $\theta\approx r/z$ at small angles.
Results of this method are presented on Fig.~\ref{Fig2} (b,c). An increase of $\theta$ accelerates the snake instability process and allows us to observe vortex creation for the same nonlinear strength as before. The snake instability causes the soliton to breakup into a pair of vortices from $v/c=\theta=5.2\times 10^{-4}$ rad. Beyond $\theta=8.7\times 10^{-4}$ rad, too many fluctuations are created making the observation of snake instability decay unclear. It is possible to approximate the speed of sound as $c_s\approx c\sqrt{\Delta n}$ which gives in our case $c_s/c \approx 9.4\times10^{-3}$. Therefore, regarding to the breakup velocity $v/c$, an additional initial velocity between $5$ to $10\%$ of $c_s$ is enough to observe vortex nucleation. This low (compared to speed of sound) velocity, which is necessary to reach a vortex generation regime, is a good indicator that the subsequent vortex dynamics occurs as in a nearly incompressible fluid. Fig.~\ref{Fig2} (d) shows the measurement of the phase gradient (i.e. fluid's velocity) close to the vortices. The velocity circulation around each intensity minimum confirms these minima contain vortices. A snake instability always leads to breaking up into a pair of oppositely charged vortices in order to conserve the total topological charge of the system to be zero. We also observe an augmentation of the fluid's velocity between the vortices. Taking into account the residual absorption, the healing length of the medium around vortices is estimated as $\Lambda\approx50\mu$m. The core diameter of the vortices (at $0.9$ of the surrounding fluid density) is around $2\Lambda=100\mu$m and the center to center distance between vortices is $100\mu$m.\\
Detailed numerical simulations indicate that the saturation and non-locality, which potentially could drastically reduce vortex nucleation, are negligible  in our regime.
Due to the geometry of our configuration, the vortices move away from each other and do not interact. Different initial conditions suitable for vortex creation need to be used in order to observe vortex interaction processes.
We therefore shift the elliptical beam along the z-axis to give the beam a curvature when entering the medium which corresponds to an additional phase gradient (fluid velocity) over several wavelengths. In this convergent configuration (radius of curvature is around $R=-0.5$m), the extra velocity pushes vortices toward the center due to its specific phase gradient distribution. This specific case leads to the creation of 4 vortex pairs visible in Fig.~\ref{Fig3} (a). Sign of topological charges is conventionally defined depending on the winding direction as follows: $+$ for anti-clockwise and $-$ for clockwise velocity. As in the previous case, each pair is composed of oppositely charged vortices (i.e. the topological charges of the vortices in the center of the structure are also opposite). Vortices move closer to each other (represented by arrows) due to the specific convergent configuration giving an initial velocity in this direction. 

Then, they collide as shown in (b) and annihilate, leading to  radiative loss. Panels (c) and (d) highlight the emergence of this loss as sound waves visible around vortices on the supplementary video. After the vortex annihilation, vortices form new pairs and the initial velocity disappears: after frame (d), the remaining vortices will move away from the center due to the non-hydrodynamic motion related to the overall beam expansion.

\section{Relative phase scanning}

\begin{figure}[!htb]
\center{
  \includegraphics[width=1\linewidth]{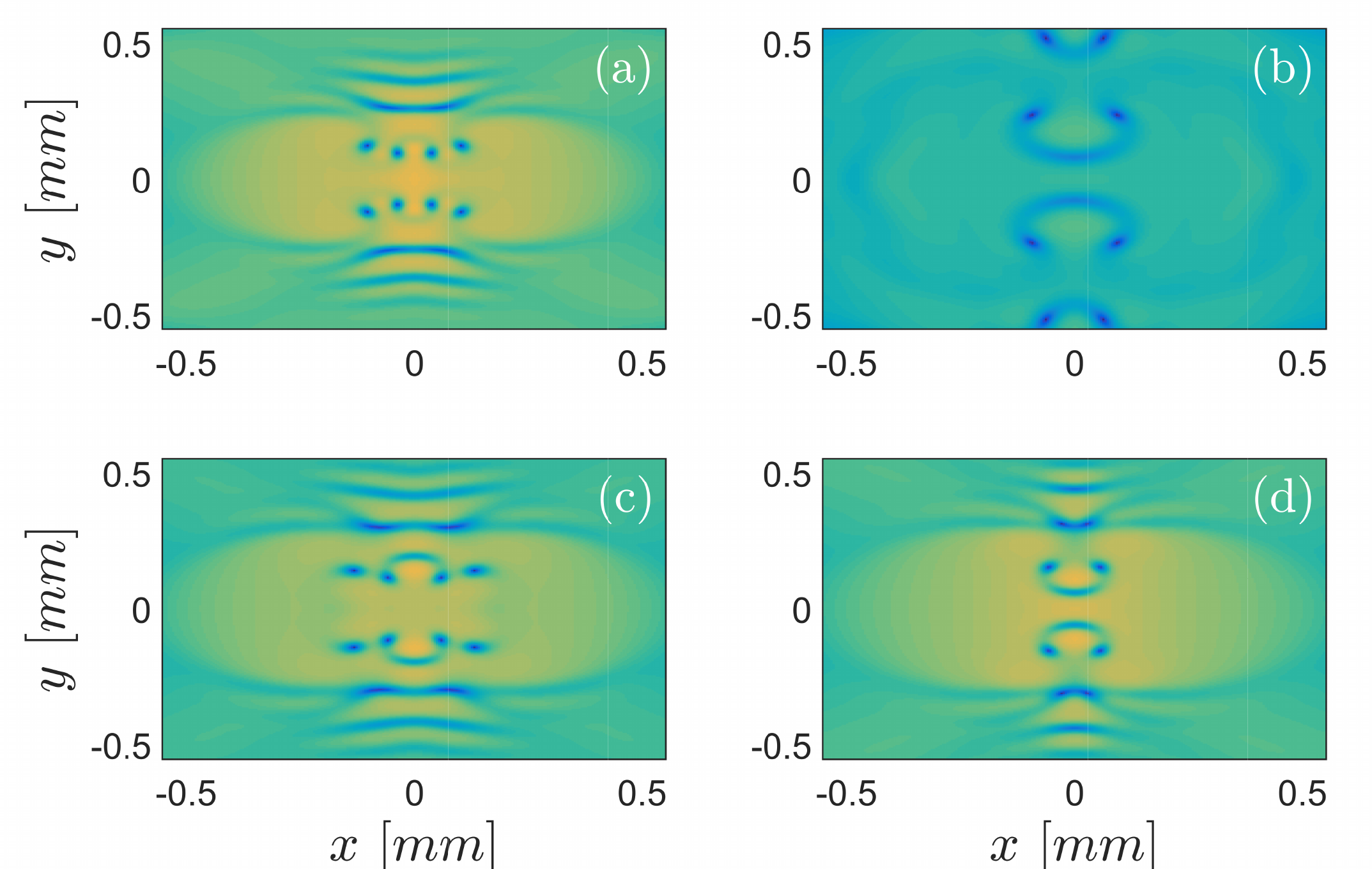}
\caption{ Numerical simulations of the converging configuration. (a) and (b) correspond to a propagation along time axis respectively $z/z_{NL}=35$ and $z/z_{NL}=70$ while (c) and (d) have the same effective time of propagation $z/z_{NL}=50$ but relative phase is respectively $0.97\pi$ and $1.05\pi$}
\label{Fig4}}
\end{figure}

The experiment presented in Fig.~\ref{Fig3} has been done at fixed nonlinearity and therefore at fixed effective time. The evolution depicted in this figure solely arises from a relative phase scanning between the elliptical and the background beam. Small dephasings ($0.9\pi<\varphi<1.1\pi$) were sufficient to observe the whole process, from the snake instabilities decaying into vortices to the vortex interactions and annihilation.
An analogy can be made between this method and  Taylor's frozen turbulence hypothesis \cite{Taylor38}. Namely, considering turbulence velocity fluctuations to be small compared to the mean flow velocity, one can approximate the time evolution of the system as its translation in the mean flow direction. In our case, scanning the relative phase changes the added velocity amplitude which remains oriented toward the center and therefore a given velocity amplitude corresponds to a specific effective time.
We observe 'time-like' dynamics of the instability and subsequent vortex motion and interaction by changing the relative phase between the two beams in the initial condition. The small change of phase is only time-like when the velocity of the fluid in the initial condition is parallel or anti-parallel to the direction of motion of the soliton, that is, perpendicular to the long axis of the elliptical beam. For a limited range of values of the relative phase, the desired direction of the velocity is maintained with only the magnitude being altered. The subsequent time-like dynamics can be understood by considering the translational motion of a 1D soliton which can be approximated to be of the form $f(x - v_s t)$. In this case $v_s$ can rescale $t$. As in 1D there is no instability, this change only affects the position of the soliton. In our case the decay of the soliton is related to the transverse density variations of the background beam through which the soliton has propagated.\\
Fig.~\ref{Fig4} presents results of numerical simulations obtained with the converging elliptical beam as an initial condition. Upper plots correspond to a usual evolution where the system evolves in time while the lower ones are equivalent to measurements shown in Fig.~\ref{Fig3}.
In addition to the good agreement between experiment and simulation when scanning the relative phase, we stress that such an agreement between the two numerical setups justifies the mapping between time evolution and phase scanning.\\
The modification of the relative phase induces small changes in the phase gradient which can be understood as a precise tuning of the initial velocity modulus. By keeping the nonlinearity fixed (therefore the evolution time) and controlling initial velocity modulus, the system will finish its propagation at different moments of the evolution of vortices. Using the time evolution only, one needs to scan the nonlinearity  from $z/z_{NL}=35$ to $70$ to observe the whole process while a phase scanning protocol only requires $z/z_{NL}=50$. Fixing the nonlinearity during the evolution process has many advantages. In particular changing $z/z_{NL}$ has an non-negligible impact on beam's dispersion.

\begin{figure}[!htb]
\centerline{
  \includegraphics[width=1\linewidth]{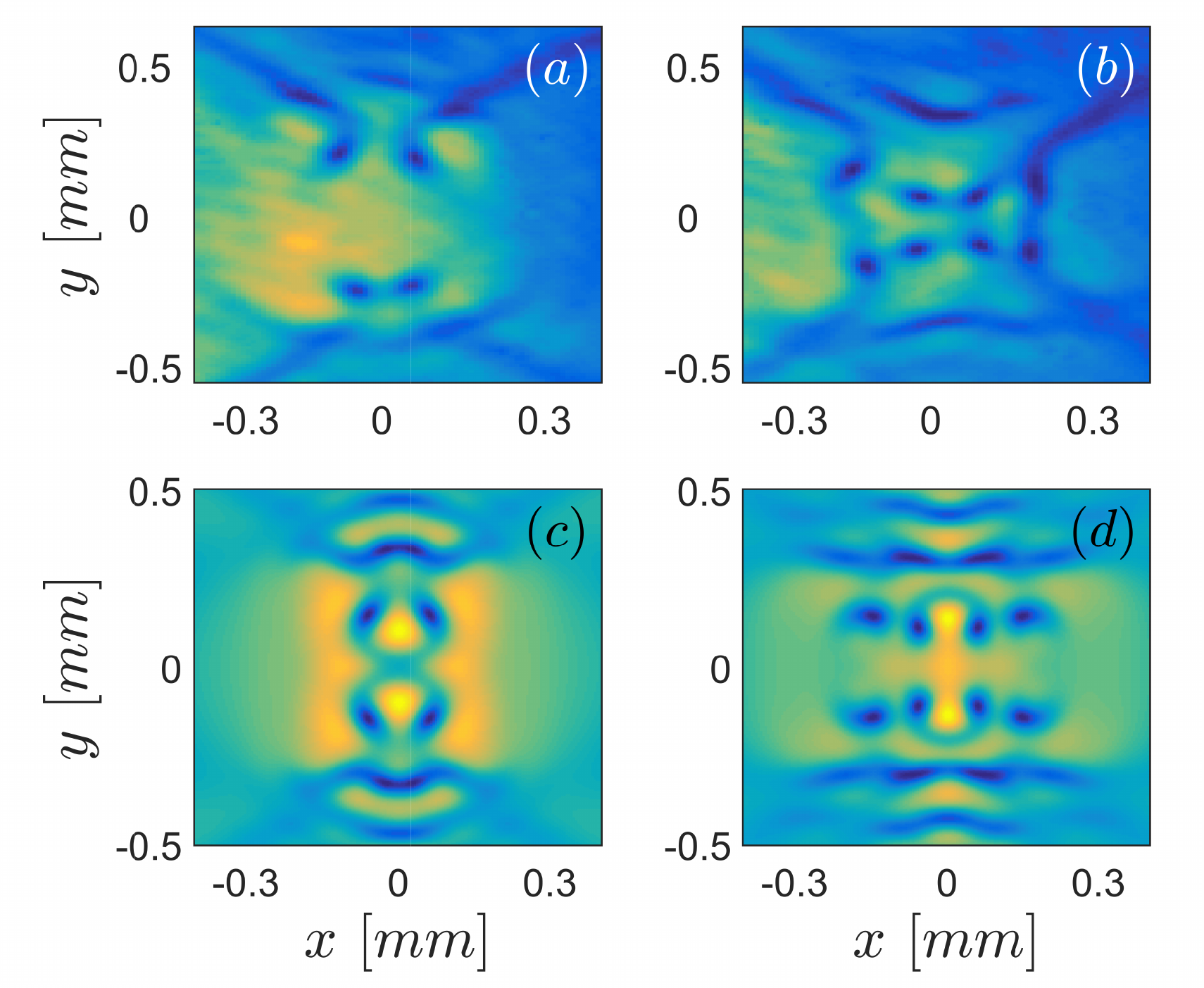}}
	\caption{Experiment (a,b) versus numerical simulation (c,d) of the diverging case $R=0.5$m. Evolution is done by scanning the relative phase: (a,c) and (b,d) correspond respectively to $\varphi=0.95\pi \text{ and } 1.08\pi$}
\label{Fig5}
\end{figure}

The mean fluid density is also reduced during the time evolution while it remains constant when scanning the relative phase: this gives the possibility to study vortex dynamics with fixed $z_{NL}$ and $\Lambda$ during the whole process. Finally, it allows us to study effective time evolution while fixing the nonlinearity at a lower value. All these points make it a powerful tool to study vortex processes.\\
Finally, we have studied the diverging configuration through experiment and numerical simulations. Oppositely to the previous case, the radius of curvature is now $R=0.5$m. The relative phase is scanned in the same direction as for the converging case presented in Fig.~\ref{Fig3} (from $\varphi<\pi$ to $\varphi>\pi$). We observe in Fig.~\ref{Fig5} that the system starts with two pairs of vortices and progressively evolves into 4 pairs of vortices. In fact, we are  in a situation where the evolution goes in the opposite direction compared to the converging case. Considering the initial phase distribution which is opposite in these two cases, the evolution of the phase gradient (initial velocity) in each case for a same relative phase is also opposite.\\ Therefore, scanning the relative phase in the same direction but in the diverging case (Fig.~\ref{Fig5}) will progressively reduce the initial velocity instead of increasing it as we saw in Fig.~\ref{Fig3}. This gives an impression of a backward process.

\section{Conclusions} 

Our new experimental techniques have allowed us to create and observe the nonlinear dynamics and annihilation of strongly interacting vortices. The vortices were created via a snake instability  of solitons arising from an initial  elliptical dark stripe. The use of a wave-front sensor has greatly simplified the vortex observation and  characterisation by allowing us to observe the wavefront in situ. Accelerating the dynamics allowed us to greatly reduce the needed nonlinearity. Specific initial conditions on the dark stripe such as giving a curvature to wavefront of the soliton leads to vortex interactions and annihilations. We  have also developed a method based on the relative phase shift between the fluid and the soliton which is equivalent to an effective time shift. This way, it is possible to observe vortex dynamics without changing any parameter of the nonlinear medium improving the interpretability of the results. Numerical simulations have confirmed these results.
The techniques developed and tested in this work are scalable to bigger experiments 
%\sout{and they can be used} 
and are aimed to produce and control systems with a greater number of vortices engaged in chaotic interactions of hydrodynamic type. 
Thus, our work gives new tools to control and study optical vortex interactions and all phenomena related to it
%\sout{is a step toward achieving}
such as fully developed 2D turbulence in optical fluids.
An initial density jump like the one in the Josephson junctions setup \cite{Griffin20} is another promising setup for creating turbulence in future experiments.
\\
This work was supported by the  EU  Horizon  2020  research and  innovation  programme  in  the  framework  of  Marie Skodowska-Curie  HALT  project  (agreement  No 823937) and the FET Flagships PhoQuS project (agreement No 820392). This work also benefited from funding of the project OPTIMAL granted by the European Union by means of the Fond Européen de Développement Régional (FEDER).

\section*{Appendix: Evaluation of the temperature of the atomic vapor}

\begin{figure}[h!]
\centerline{{\includegraphics[width=1 \linewidth]{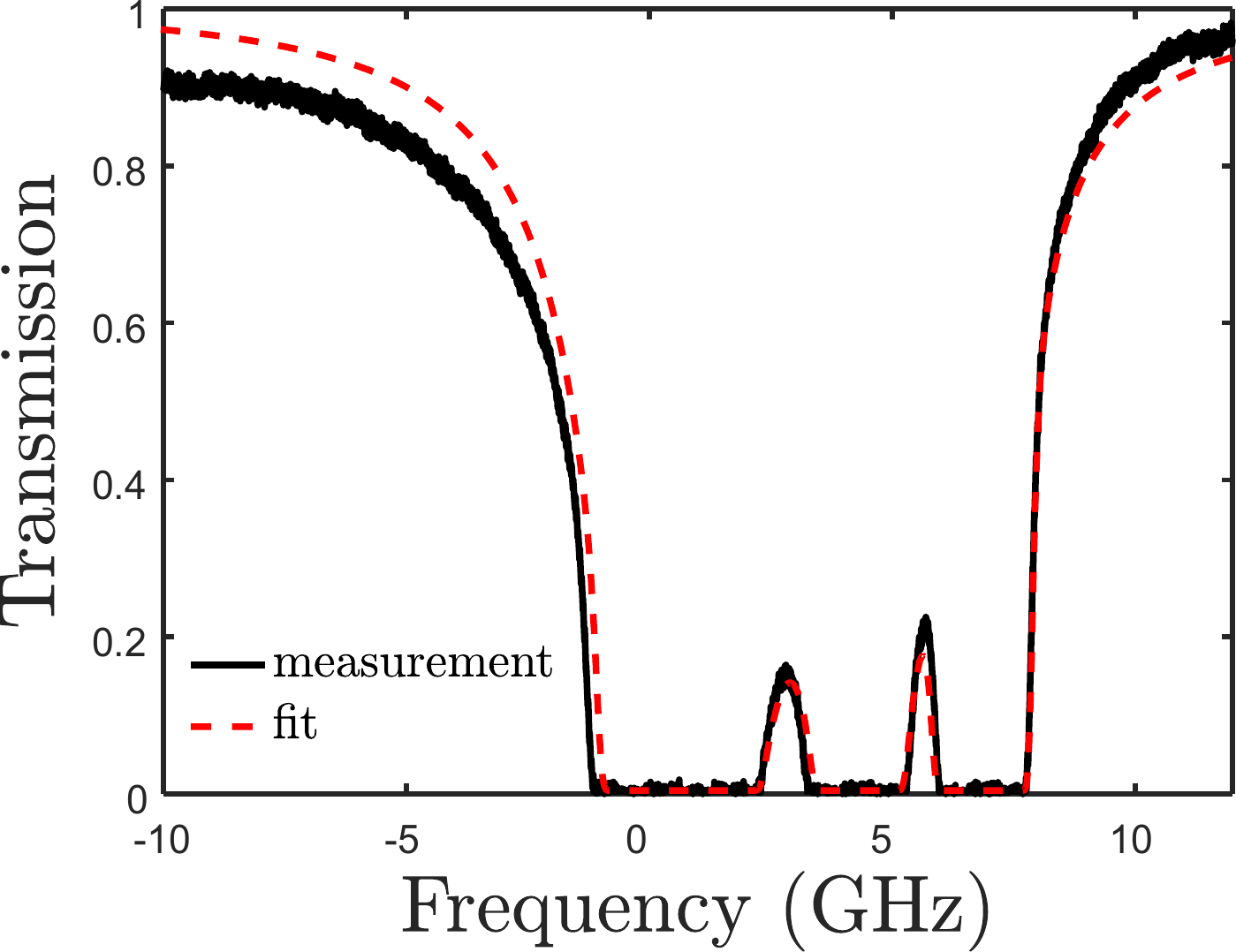}}}
	\caption{Transmission of the laser beam through the cell as a function of the detuning $\delta$ of the laser. From the transmission profile a fit to the data allows us to deduce the temperature of the gas. For this particular curve, we find $T=110^o$C.}
\label{Fig6}
\end{figure}

To extract atomic gas temperature, we measure the transmission profile of a weak laser beam after propagation through a the vapor as a function of the laser detuning. Fitting this data (as shown in Fig.\ref{Fig6}) with a numerical simulation taking into account atomic lines of both isotopes, rubidium vapor pressure as a function of the temperature \cite{siddons2008absolute,agha2011time,Baudouin14} and the Doppler broadening, we can deduce the atomic density and therefore link it to the gas temperature using the vapor pressure \cite{steck} and ideal gas law.\\

%\bibliography{biblio_vortex}


\end{document}